\pdfoutput=1

\documentclass[12pt,a4paper]{article}
\usepackage{ifthen} 
\newboolean{pdflatex}
\setboolean{pdflatex}{true} 
\newboolean{articletitles}
\setboolean{articletitles}{true} 
\newboolean{uprightparticles}
\setboolean{uprightparticles}{false} 
\newboolean{inbibliography}
\setboolean{inbibliography}{false} 

\usepackage{microtype}
\usepackage{lineno}  
\usepackage{xspace} 
\usepackage{caption}

\usepackage{graphicx} 
\usepackage{color}
\usepackage{colortbl}
\graphicspath{{./figs/}} 
\usepackage{amsmath} 
\usepackage{amssymb}
\usepackage{amsfonts}
\usepackage{upgreek} 
\usepackage{hyperref}  
\usepackage[all]{hypcap} 
\usepackage{xspace} 
\usepackage{upgreek}

\def\lhcb {\mbox{LHCb}\xspace}

\ifthenelse{\boolean{uprightparticles}}
{

 \def\Ppi         {\ensuremath{\uppi}\xspace}

 \def\PDelta      {\ensuremath{\Delta}\xspace}                 
 \def\PXi      {\ensuremath{\Xi}\xspace}                 
 \def\PLambda      {\ensuremath{\Lambda}\xspace}                 
 \def\PSigma      {\ensuremath{\Sigma}\xspace}                 
 \def\POmega      {\ensuremath{\Omega}\xspace}                 
 \def\PUpsilon      {\ensuremath{\Upsilon}\xspace}                 
                  
 \def\PB      {\ensuremath{\mathrm{B}}\xspace}                 
                  
 \def\PD      {\ensuremath{\mathrm{D}}\xspace}

 \def\PK      {\ensuremath{\mathrm{K}}\xspace}

 \def\Pb      {\ensuremath{\mathrm{b}}\xspace}                 
 \def\Pc      {\ensuremath{\mathrm{c}}\xspace}                 
                  
 \def\Pe      {\ensuremath{\mathrm{e}}\xspace}

 \def\Pi      {\ensuremath{\mathrm{i}}\xspace}

}
{

 \def\Ppi         {\ensuremath{\pi}\xspace}

 \mathchardef\PDelta="7101
 \mathchardef\PXi="7104
 \mathchardef\PLambda="7103
 \mathchardef\PSigma="7106
 \mathchardef\POmega="710A
 \mathchardef\PUpsilon="7107
                  
 \def\PB      {\ensuremath{B}\xspace}                 
                  
 \def\PD      {\ensuremath{D}\xspace}

 \def\PK      {\ensuremath{K}\xspace}

 \def\Pb      {\ensuremath{b}\xspace}                 
 \def\Pc      {\ensuremath{c}\xspace}                 
                  
 \def\Pe      {\ensuremath{e}\xspace}

 \def\Pi      {\ensuremath{i}\xspace}

}

\makeatletter
\ifcase \@ptsize \relax
  \newcommand{\miniscule}{\@setfontsize\miniscule{4}{5}}
\or
  \newcommand{\miniscule}{\@setfontsize\miniscule{5}{6}}
\or
  \newcommand{\miniscule}{\@setfontsize\miniscule{5}{6}}
\fi
\makeatother

\DeclareRobustCommand{\optbar}[1]{\shortstack{{\miniscule (\rule[.5ex]{1.25em}{.18mm})}
  \\ [-.7ex] $#1$}}

\def\epem       {{\ensuremath{\Pe^+\Pe^-}}\xspace}

\def\cquark    {{\ensuremath{\Pc}}\xspace}

\def\bquark    {{\ensuremath{\Pb}}\xspace}

\def\pion   {{\ensuremath{\Ppi}}\xspace}

\def\pip    {{\ensuremath{\pion^+}}\xspace}
\def\pim    {{\ensuremath{\pion^-}}\xspace}
\def\pipm   {{\ensuremath{\pion^\pm}}\xspace}
\def\pimp   {{\ensuremath{\pion^\mp}}\xspace}

\def\kaon    {{\ensuremath{\PK}}\xspace}
\def\Kbar    {{\kern 0.2em\overline{\kern -0.2em \PK}{}}\xspace}

\def\KorKbar    {\kern 0.18em\optbar{\kern -0.18em K}{}\xspace}

\def\Kp      {{\ensuremath{\kaon^+}}\xspace}
\def\Km      {{\ensuremath{\kaon^-}}\xspace}
\def\Kpm     {{\ensuremath{\kaon^\pm}}\xspace}
\def\Kmp     {{\ensuremath{\kaon^\mp}}\xspace}
\def\KS      {{\ensuremath{\kaon^0_{\mathrm{ \scriptscriptstyle S}}}}\xspace}

\def\Dbar    {{\kern 0.2em\overline{\kern -0.2em \PD}{}}\xspace}
\def\D       {{\ensuremath{\PD}}\xspace}

\def\DorDbar    {\kern 0.18em\optbar{\kern -0.18em D}{}\xspace}
\def\Dz      {{\ensuremath{\D^0}}\xspace}
\def\Dzb     {{\ensuremath{\Dbar{}^0}}\xspace}

\def\Dstarp  {{\ensuremath{\D^{*+}}}\xspace}
\def\Dstarm  {{\ensuremath{\D^{*-}}}\xspace}
\def\Dstarpm {{\ensuremath{\D^{*\pm}}}\xspace}

\def\B       {{\ensuremath{\PB}}\xspace}
\def\Bbar    {{\ensuremath{\kern 0.18em\overline{\kern -0.18em \PB}{}}}\xspace}

\def\BorBbar    {\kern 0.18em\optbar{\kern -0.18em B}{}\xspace}

\def\Bu      {{\ensuremath{\B^+}}\xspace}

\def\Bp      {{\ensuremath{\Bu}}\xspace}

\def\Bpm     {{\ensuremath{\B^\pm}}\xspace}

  \def\Y#1S{\ensuremath{\PUpsilon{(#1S)}}\xspace}

\def\Lbar        {{\ensuremath{\kern 0.1em\overline{\kern -0.1em\PLambda}}}\xspace}
\def\LorLbar    {\kern 0.18em\optbar{\kern -0.18em \PLambda}{}\xspace}

\newcommand{\decay}[2]{\ensuremath{#1\!\to #2}\xspace}         

\def\to                 {\ensuremath{\rightarrow}\xspace}

\def\CP                {{\ensuremath{C\!P}}\xspace}

\def\AT#1     {\ensuremath{A_{\mathrm{T}}^{#1}}\xspace}           % 2

\def\C#1      {\ensuremath{\mathcal{C}_{#1}}\xspace}                       % 9
\def\Cp#1     {\ensuremath{\mathcal{C}_{#1}^{'}}\xspace}                    % 7
\def\Ceff#1   {\ensuremath{\mathcal{C}_{#1}^{\mathrm{(eff)}}}\xspace}        % 9  
\def\Cpeff#1  {\ensuremath{\mathcal{C}_{#1}^{'\mathrm{(eff)}}}\xspace}       % 7
\def\Ope#1    {\ensuremath{\mathcal{O}_{#1}}\xspace}                       % 2
\def\Opep#1   {\ensuremath{\mathcal{O}_{#1}^{'}}\xspace}                    % 7

\newcommand{\ket}[1]{\ensuremath{|#1\rangle}}              % {b}
\newcommand{\tev}{\ifthenelse{\boolean{inbibliography}}{\ensuremath{~T\kern -0.05em eV}\xspace}{\ensuremath{\mathrm{\,Te\kern -0.1em V}}}\xspace}
\newcommand{\gev}{\ensuremath{\mathrm{\,Ge\kern -0.1em V}}\xspace}
\newcommand{\mev}{\ensuremath{\mathrm{\,Me\kern -0.1em V}}\xspace}
\newcommand{\kev}{\ensuremath{\mathrm{\,ke\kern -0.1em V}}\xspace}
\newcommand{\ev}{\ensuremath{\mathrm{\,e\kern -0.1em V}}\xspace}
\newcommand{\gevc}{\ensuremath{{\mathrm{\,Ge\kern -0.1em V\!/}c}}\xspace}
\newcommand{\mevc}{\ensuremath{{\mathrm{\,Me\kern -0.1em V\!/}c}}\xspace}
\newcommand{\gevcc}{\ensuremath{{\mathrm{\,Ge\kern -0.1em V\!/}c^2}}\xspace}
\newcommand{\gevgevcccc}{\ensuremath{{\mathrm{\,Ge\kern -0.1em V^2\!/}c^4}}\xspace}
\newcommand{\mevcc}{\ensuremath{{\mathrm{\,Me\kern -0.1em V\!/}c^2}}\xspace}

\def\invfb   {\ensuremath{\mbox{\,fb}^{-1}}\xspace}

\newcommand{\chisq}{\ensuremath{\chi^2}\xspace}

\def\gsim{{~\raise.15em\hbox{$>$}\kern-.85em
          \lower.35em\hbox{$\sim$}~}\xspace}
\def\lsim{{~\raise.15em\hbox{$<$}\kern-.85em
          \lower.35em\hbox{$\sim$}~}\xspace}

\def\ptot       {\mbox{$p$}\xspace}
\def\pt         {\mbox{$p_{\mathrm{ T}}$}\xspace}

\newcommand{\ie}{\mbox{\itshape i.e.}\xspace}

\newcommand{\cleoc}{\mbox{{C}{L}{E}{O}{-}{c}}\xspace}

\def\DzTof       {\decay{\Dz}{f}}
\def\DzbTof       {\decay{\Dzb}{f}}

\def\DzToKmpippimpip       {\decay{\Dz}{\Km\pip\pim\pip}}

\def\DzToKppimpippim       {\decay{\Dz}{\Kp\pim\pip\pim}}

\def\DToKmppipmpimppipm       {\decay{\D}{\Kmp\pipm\pimp\pipm}}

\def\DToKmpippimpip       {\decay{\D}{\Km\pip\pim\pip}}
\def\DzToKpipipiWS      {\DzToKppimpippim}
\def\DzToKpipipiRS       {\DzToKmpippimpip}

\def\DstarpFull  {{\ensuremath{\D^{*}(2010)^{+}}}\xspace}
\def\DstarmFull  {{\ensuremath{\D^{*}(2010)^{-}}}\xspace}

\def\slowpion                  {{\ensuremath{\pi_s}}\xspace}
\def\slowpionp                  {{\ensuremath{\pi_s^{+}}}\xspace}
\def\slowpionm                  {{\ensuremath{\pi_s^{-}}}\xspace}
\def\DstarpToDzpip         {\decay{\DstarpFull}{\Dz\slowpionp}}
\def\DstarmToDzbpim      {\decay{\DstarmFull}{\Dzb\slowpionm}}
\def\DzTag                       {\DstarpToDzpip}
\def\DzbTag                     {\DstarmToDzbpim}

\def\DDb                        {{\ensuremath{\D\Dbar}}\xspace}
\def\DzDzb                     {{\ensuremath{\Dz}--\ensuremath{\Dzb}}\xspace}

\def\BpToDKp          {\decay{\Bp}{\D\Kp}}
\def\BpmToDKpm    {\decay{\Bpm}{\D\Kpm}}

\newcommand{\eqnref}[1]{Eq.~\ref{#1}}

\newcommand{\Figref}[1]{Figure~\ref{#1}}
\newcommand{\figref}[1]{Fig.~\ref{#1}}

\newcommand{\tabref}[1]{Table~\ref{#1}}

\newcommand{\RKpipipi}{\ensuremath{R^{K3\pi}_{D}}\xspace}
\newcommand{\delKpipipi}{\ensuremath{\delta_{D}^{K3\pi}}\xspace}
\newcommand{\rKpipipi}{\ensuremath{r^{K3\pi}_{D}}\xspace}

\newcommand{\yprimekpipipi}{\ensuremath{y'_{K3\pi}}\xspace}

\newcommand{\gam}{\ensuremath{\gamma}\xspace}

\newcommand{\deltam}{\ensuremath{\Delta m}\xspace}

\def\Dzket {\ket{\Dz}\xspace} 
\def\Dzbket {\ket{\Dzb}\xspace}

\def\chisq {\ensuremath{\chi^{2}}\xspace}

\newcommand{\seccor}{\ensuremath{\Delta_{\mathrm{sec},i}}\xspace}
\newcommand{\seccorj}{\ensuremath{\Delta_{\mathrm{sec},j}}\xspace}
\newcommand{\secfrac}{\ensuremath{f_{\mathrm{sec},i}}\xspace}
\newcommand{\kscor}{\ensuremath{\Delta_{\KS}}\xspace}
\newcommand{\idcori}{\ensuremath{\Delta_{\mathrm{ID},i}}\xspace}
\newcommand{\effcori}{\ensuremath{\epsilon_{i}}\xspace}

\usepackage{cite} % Allows for ranges in citations
\usepackage{mciteplus}
\usepackage{longtable} % only for template; not usually to be used in PAPERs

\begin{document}

\renewcommand{\thefootnote}{\fnsymbol{footnote}}
\setcounter{footnote}{1}

%%%%%%%%%%%%%%%%%%%%%%%%%
%%%%%  TITLE PAGE  %%%%%%
%%%%%%%%%%%%%%%%%%%%%%%%%
\begin{titlepage}
\pagenumbering{roman}

% Header ---------------------------------------------------
\vspace*{-1.5cm}
\centerline{\large EUROPEAN ORGANIZATION FOR NUCLEAR RESEARCH (CERN)}
\vspace*{1.5cm}
\noindent
\begin{tabular*}{\linewidth}{lc@{\extracolsep{\fill}}r@{\extracolsep{0pt}}}
\ifthenelse{\boolean{pdflatex}}
{\vspace*{-2.7cm}\mbox{\!\!\!\includegraphics[width=.14\textwidth]{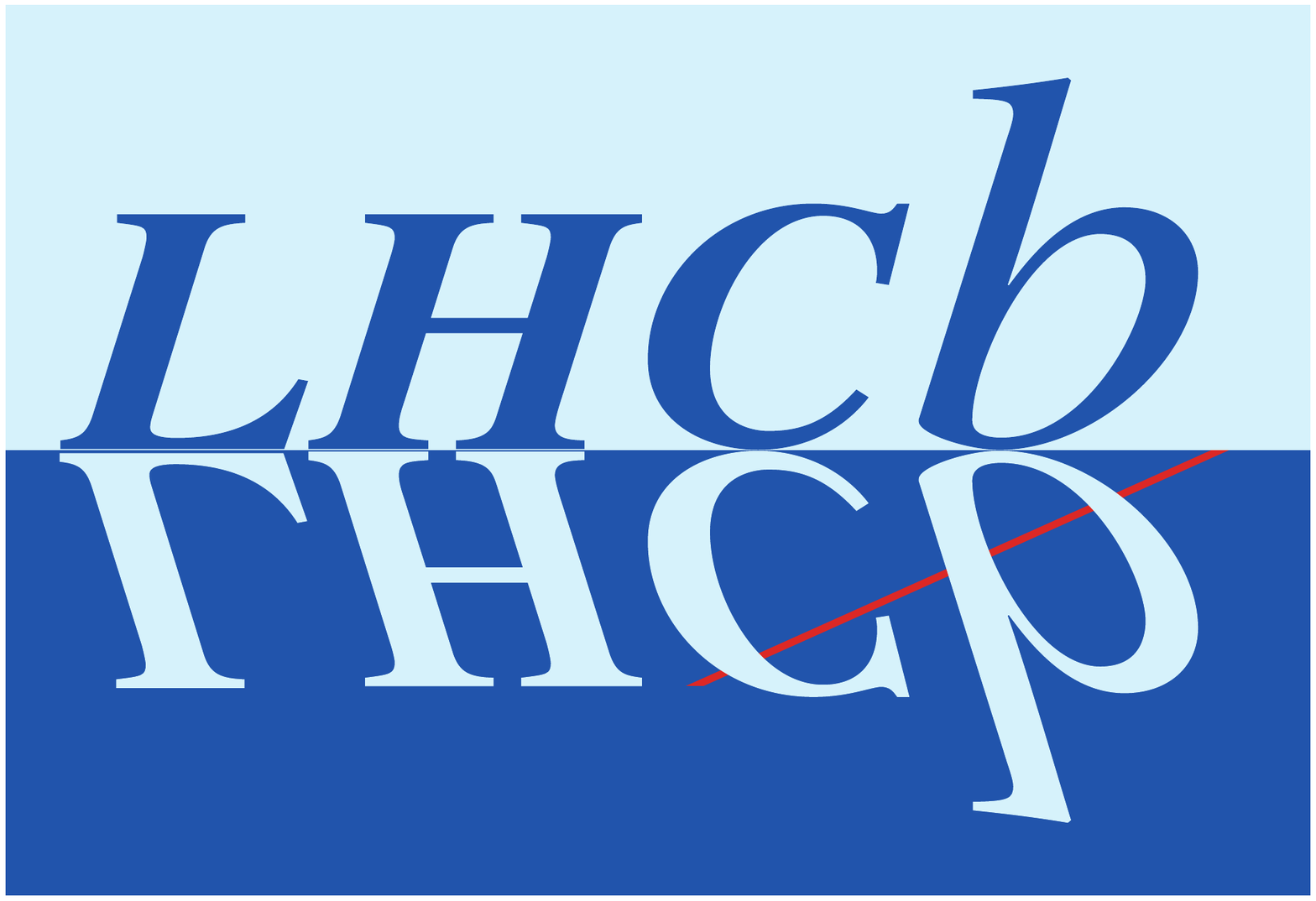}} & &}%
{\vspace*{-1.2cm}\mbox{\!\!\!\includegraphics[width=.12\textwidth]{lhcb-logo.eps}} & &}%
\\
 & & CERN-EP-2016-021 \\  
 & & LHCb-PAPER-2015-057 \\  
 & & 17 June 2016 \\ 
 & & \\
\end{tabular*}

\vspace*{2.0cm}

{\normalfont\bfseries\boldmath\huge
\begin{center}
First observation of \DzDzb oscillations in $\Dz\to\Kp\pim\pip\pim$ decays and measurement of the associated coherence parameters
%% Simplified for arXiv submission:  
%% First observation of $D^0-\bar D^0$ oscillations in $D^0\to K^+\pi^-\pi^+\pi^-$ decays and measurement of the associated coherence parameters   
 \end{center}
}

\vspace*{0.2cm}

\begin{center}
The LHCb collaboration\footnote{Authors are listed at the end of this Letter.}
\end{center}

\vspace{\fill}

\begin{abstract}

  \noindent Charm meson oscillations are observed in a time-dependent analysis
  of the ratio of $\Dz\to\Kp\pim\pip\pim$ to $\Dz\to\Km\pip\pim\pip$
  decay rates, using data corresponding to an integrated luminosity of
  3.0\invfb recorded by the \lhcb experiment. 
  The measurements presented are sensitive to the phase-space averaged ratio of
  doubly Cabibbo-suppressed to Cabibbo-favoured amplitudes \rKpipipi and 
  the product of the coherence factor
  \RKpipipi\ and a charm mixing parameter \yprimekpipipi.
  The constraints measured are $\rKpipipi=(5.67 \pm 0.12)\times10^{-2}$, which is the most precise determination to date, and $\RKpipipi \cdot \yprimekpipipi = (0.3 \pm 1.8)\times 10^{-3}$, which provides useful input for determinations of the CP-violating phase \gam in \BpmToDKpm, \DToKmppipmpimppipm decays.  The analysis also gives the most precise measurement of the
  $\Dz\to\Kp\pim\pip\pim$ branching fraction, and the first
  observation of \DzDzb oscillations in this decay mode, with a significance of 8.2 standard deviations.

\end{abstract}

%%  Simplified for arXiv submission:  
%%  \noindent Charm meson oscillations are observed in a time-dependent analysis of the ratio of $D^0\to K^+\pi^-\pi^+\pi^-$ to $D^0\to K^-\pi^+\pi^-\pi^+$ decay rates, using data corresponding to an integrated luminosity of $3.0\,{\rm fb}^{-1}$ recorded by the LHCb experiment. The measurements presented are sensitive to the phase-space averaged ratio of doubly Cabibbo-suppressed to Cabibbo-favoured amplitudes $r_{D}^{K3\pi}$ and the product of the coherence factor $R_{D}^{K3\pi}$ and a charm mixing parameter $y^{'}_{K3\pi}$. The constraints measured are $r_{D}^{K3\pi}=(5.67 \pm 0.12)\times10^{-2}$, which is the most precise determination to date, and $R_{D}^{K3\pi} \cdot y^{'}_{K3\pi} = (0.3 \pm 1.8)\times 10^{-3}$, which provides useful input for determinations of the CP-violating phase $\gamma$ in $B^\pm \to D K^\pm, D \to K^\mp\pi^\pm\pi^\mp\pi^\pm$ decays.  The analysis also gives the most precise measurement of the $D^0\to K^+\pi^-\pi^+\pi^-$ branching fraction, and the first observation of $D^0-\bar D^0$ oscillations in this decay mode, with a significance of 8.2 standard deviations.

\vspace*{1.0cm}

\begin{center}
  Published in Phys.~Rev.~Lett.~116, 241801 (2016)
\end{center}

\vspace{\fill}

{\footnotesize 
\centerline{\copyright~CERN on behalf of the \lhcb collaboration, licence \href{http://creativecommons.org/licenses/by/4.0/}{CC-BY-4.0}.}}
\vspace*{2mm}

\end{titlepage}

\newpage
\setcounter{page}{2}
\mbox{~}

\cleardoublepage

\renewcommand{\thefootnote}{\arabic{footnote}}
\setcounter{footnote}{0}

%%%%%%%%%%%%%%%%%%%%%%%%%
%%%%% Main text %%%%%%%%%
%%%%%%%%%%%%%%%%%%%%%%%%%

\pagestyle{plain} 
\setcounter{page}{1}
\pagenumbering{arabic}

%%%%%%%%%%%%%%%%%%%%%%%%%
%%%%% 1 %%%%%%%%%
%%%%%%%%%%%%%%%%%%%%%%%%%

\noindent Neutral mesons can oscillate between their particle and
anti-particle states. 
This phenomenon, also referred to as
mixing, is of considerable interest for a variety of reasons,
including its unique sensitivity to effects beyond the Standard Model
(SM) of particle physics. Mixing has been observed in strange, beauty,
and, most recently, charm mesons. Its observation in the charm ($\Dz -
\Dzb$) system is particularly challenging, with an oscillation period
that is more than $1000$ times longer than the meson's lifetime. It
took until 2008 for charm mixing to be established, by combining
results from BaBar, BELLE and CDF~\cite{BaBar:Mixing2007,
Belle:Mixing2007, CDF:Mixing2008, HFAG, *HFAG2015}, and until 2013 for
the first $5\sigma$ observation in an individual
measurement~\cite{LHCb-PAPER-2012-038}. Until now, all $5\sigma$
observations of charm mixing in individual measurements have been made
in the decay mode $\Dz \to \Kp\pim$~\cite{LHCb-PAPER-2012-038,
CDF2013:Mixing, BELLE2014:Mixing}.\footnote{Unless otherwise stated,
the inclusion of charge-conjugate modes is implied throughout.} This
Letter reports the first observation of charm mixing in a different
decay channel, $\Dz\to\Kp\pim\pip\pim$. 
Previous studies of this decay mode have been consistent with the no-mixing
hypothesis~\cite{BelleK3piUpdate, Aubert:2006rz}.
Charm mixing is also sensitive to the phase difference between
charm and anti-charm decay amplitudes to the same final state.  This
phase information plays an important role in the measurement of the
charge-parity (\CP) violating phase \gam\ (or $\phi_3$), which is
accessible in decays with $b \to u$ quark transitions. 
The precision measurement of the relative magnitudes and phases 
of quark transitions provides a stringent test of the SM, and the parameter \gam
plays a central role in this effort. Currently, \gam\ has a relatively large experimental
uncertainty, and can be measured, with negligible uncertainty from theory input, in
the decay $\BpToDKp$ (and others) where \D represents a
superposition of \Dz and \Dzb states~\cite{GLW1,GLW2, ADS, GGSZ,
Rademacker:2006zx}. 
In order to constrain \gam using these decay modes, external input is
required to describe both the interference and relative magnitude of
\DzTof and \DzbTof amplitudes, where $f$ represents the final state of
the \D\ decay. Previously, it was thought that the relevant phase
information could only be measured
at \epem colliders operating at the charm threshold, where
correlated \DDb pairs provide well-defined superpositions of \Dz and
\Dzb states.  Recent studies~\cite{selfcite, selfcite2} have shown
that this input can also be obtained from a time-dependent measurement
of \DzDzb oscillations. This is the approach followed here.

%%%%%%%%%%%%%%%%%%%%%%%%%
%%%%% 2 %%%%%%%%%
%%%%%%%%%%%%%%%%%%%%%%%%%

In this work the observation of \DzDzb oscillations is made by measuring the time-dependent ratio of \DzToKpipipiWS to \DzToKpipipiRS decay rates.
The flavour of the \D meson at production is determined using the decays \DzTag and \DzbTag, where the charge of the soft (low-momentum) pion, \slowpion, tags the flavour of the meson.
The wrong-sign (WS) decay \DzToKpipipiWS has two dominant contributions: a doubly Cabibbo-suppressed (DCS) amplitude, and a \DzDzb oscillation followed by a Cabibbo-favoured (CF) amplitude.
The right-sign (RS) decay \DzToKpipipiRS is dominated by the CF amplitude, and has negligible contributions of $\mathcal{O}(10^{-4})$ from \DzDzb oscillations.
Ignoring CP violation, to second order in $t/\tau$, the time-dependence of the phase-space integrated decay rate ratio $R(t)$ is approximated by
\begin{align}
R(t) = \frac{\Gamma [ \DzToKpipipiWS ](t) }{ \Gamma[ \DzToKpipipiRS ](t) } \approx \left ( \rKpipipi \right )^{2} - \rKpipipi \RKpipipi\cdot\yprimekpipipi \frac{t}{\tau} + \frac{x^{2} + y^{2}}{4} \left ( \frac{t}{\tau} \right )^{2},
\label{eqn:ratio}
\end{align}
where $\Gamma$ denotes the decay rate, $t$ is the proper decay-time of the \Dz meson (measured with respect to production), 
$\tau$ is the \Dz lifetime, and \rKpipipi gives the phase space averaged ratio of DCS to CF amplitudes~\cite{selfcite,selfcite2}.
The dimensionless parameters $x$ and $y$ describe mixing in the \Dz meson system, with
$x$ proportional to the mass difference of the two mass eigenstates, and $y$ proportional to the width difference~\cite{HFAG}. 
Here, $\yprimekpipipi$ is defined by $\yprimekpipipi \equiv y \cos \delKpipipi - x \sin \delKpipipi$, where $\delKpipipi$ is the average strong phase difference; this and the coherence factor, $\RKpipipi$, are defined by ${\RKpipipi e^{-i \delKpipipi} \equiv \langle\cos \delta \rangle + i\langle\sin \delta \rangle}$, where ${\langle\cos \delta \rangle}$ and ${\langle\sin \delta \rangle}$ are the cosine and sine of the phase of the ratio of the DCS to the CF amplitude, averaged over phase space.\footnote{The convention $\CP \Dzket = + \Dzbket$ is followed, which determines the sign of the linear term in \eqnref{eqn:ratio}.}
For the range of \Dz decay-times used in this analysis, $[0.5,12.0]\times\tau$, \eqnref{eqn:ratio} is correct to within $\mathcal{O}(10^{-6})$.
All three parameters, \rKpipipi, \RKpipipi and \delKpipipi are required to determine \gam in \BpToDKp, \DToKmpippimpip decays.

%%%%%%%%%%%%%%%%%%%%%%%%%
%%%%% 3 %%%%%%%%%
%%%%%%%%%%%%%%%%%%%%%%%%%

This analysis is based on data samples collected in 2011 and 2012 with the \lhcb detector at centre-of-mass collision energies of $\sqrt{s} = 7\tev$ and $8\tev$ corresponding to integrated luminosities of 1.0\invfb and 2.0\invfb, respectively. 
The \lhcb detector~\cite{Alves:2008zz,LHCb-DP-2014-002} is a single-arm forward
spectrometer covering the \mbox{pseudorapidity} range $2<\eta <5$,
designed for the study of particles containing \bquark or \cquark
quarks. 
The detector elements that are particularly relevant to this analysis are: 
a silicon-strip vertex detector surrounding the $pp$ interaction region that allows \cquark- and \bquark-hadrons to be identified from their characteristically long flight distance;
a tracking system that provides a measurement of momentum, \ptot, of charged particles;
and two ring-imaging Cherenkov detectors that are able to discriminate between different species of charged hadrons.
Simulated events are produced using the software described in Refs.~\cite{Sjostrand:2006za, *Sjostrand:2007gs, LHCb-PROC-2010-056, Allison:2006ve, LHCb-PROC-2011-006}. Differences between data and simulation are corrected using data-driven techniques described in~\cite{LHCb-DP-2013-002, LHCB-PAPER-2015-028}.

%%%%%%%%%%%%%%%%%%%%%%%%%
%%%%% 4 %%%%%%%%%
%%%%%%%%%%%%%%%%%%%%%%%%%

Events are first selected by the LHCb trigger~\cite{LHCb-DP-2012-004},
and then by additional offline requirements.
Four tracks in the event must be consistent with the decay
\DzToKpipipiWS, each with momentum $\ptot >3\gevc$ and
transverse momentum $\pt > 350\mevc$.
The \Dz daughters are required to be inconsistent with originating
from a primary $pp$ interaction vertex (PV) and are combined to form a \Dz candidate, which must have a good vertex quality and $\pt>4.7\gevc$.
The soft pion, which is combined with the \Dz candidate to form a \Dstarp candidate, is required to satisfy $\ptot >3\gevc$ and $\pt > 360\mevc$.
The \Dstarp candidate must have a good vertex quality, and is reconstructed under the constraint that it originates from its associated PV.
In order to suppress backgrounds where tracks are misidentified or mis-reconstructed, information from the particle identification and tracking systems is used.
Secondary decays, \ie\ \Dstarp mesons from the decay of a $b$-hadron,
are rejected by requiring that the \Dz meson candidate is consistent with originating from a PV.
Only \Dz candidates that are reconstructed within $24\mevcc$ of the \Dz meson mass~\cite{PDG2014} are used in the analysis, reducing the amount of partially reconstructed and misidentified background. 
To reduce combinatorial background from randomly associated soft pions there is also a requirement that the invariant mass difference $\deltam \equiv m(\Kp\pim\pip\pim\pi^{\pm}_s) - m(\Kp\pim\pip\pim)$ is less than 155\mevcc.
Approximately $4\%$ of events that pass the selection requirements contain multiple signal candidates. 
In such cases one candidate is picked at random and the rest are discarded.

\begin{figure}[tb]
  \begin{center}
    \includegraphics[width=0.49\linewidth]{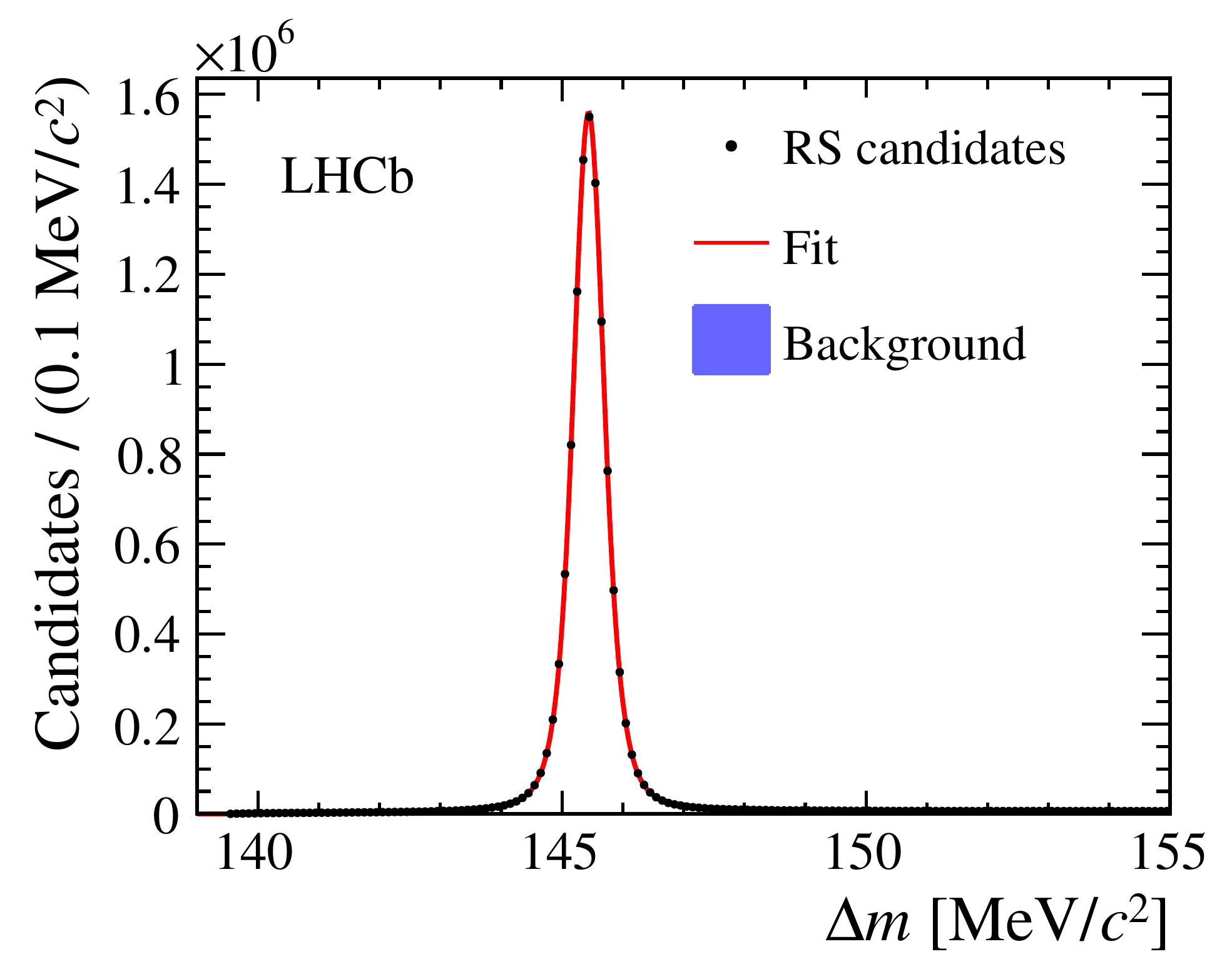}
    \includegraphics[width=0.49\linewidth]{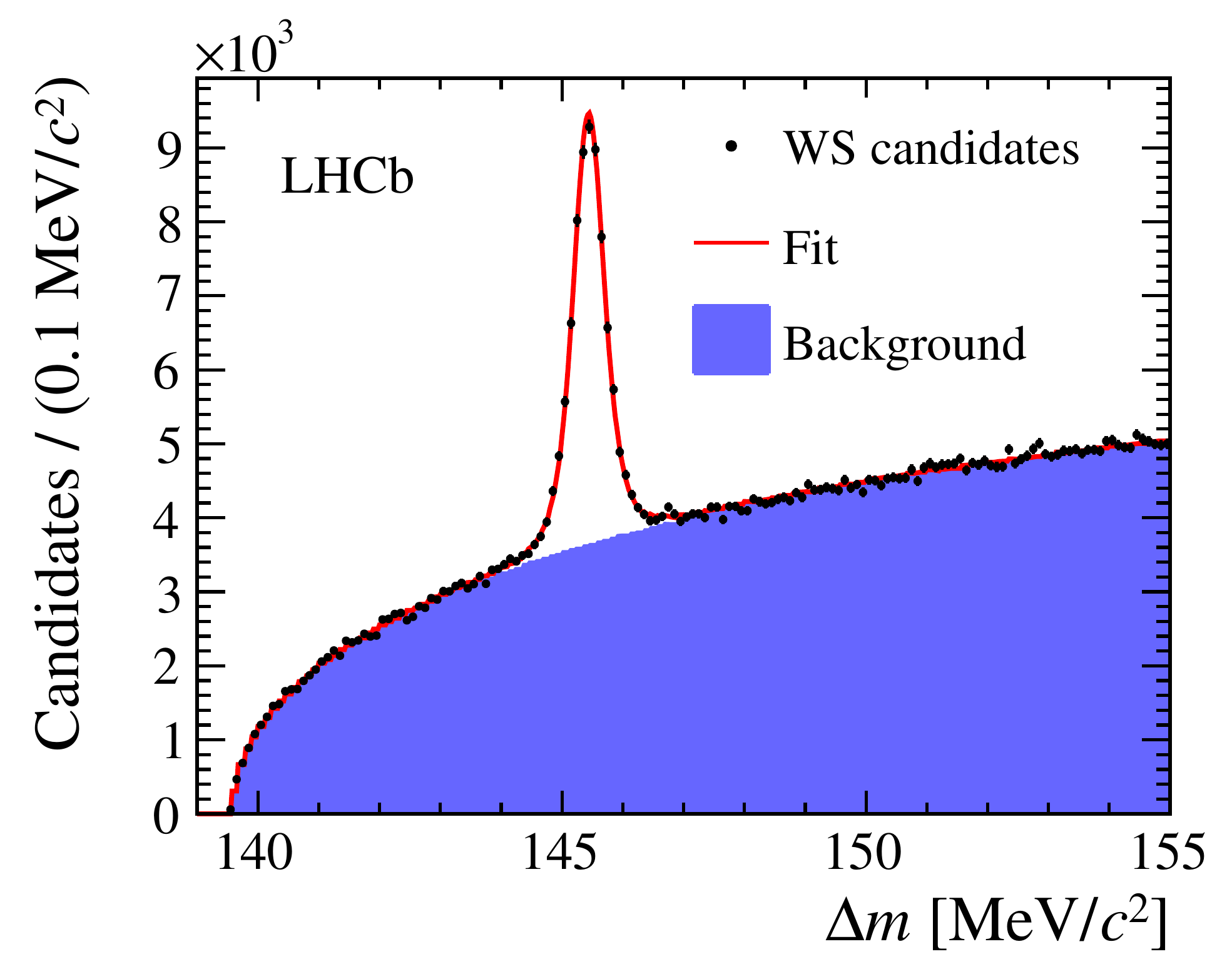}
    \vspace*{-0.7cm}
  \end{center}
  \caption{ Decay-time integrated \deltam distributions for RS (left) and WS (right) candidates with the fit result superimposed. }
  \label{fig:deltamfits}
\end{figure}

\Figref{fig:deltamfits} shows the \deltam distribution of WS and RS signal candidates with the results of a binned likelihood fit superimposed. 
The fit includes both a signal and a combinatorial background component: the signal component is empirically described by the sum of a Johnson function~\cite{JohnsonFunction} and three Gaussian functions.
The background component is estimated by randomly associating \Dz candidates with soft pions from different events. 
The resulting shape is multiplied by a first-order polynomial whose parameters are free to vary in the fit. 
The fit is made simultaneously to four decay categories: WS and RS modes for \Dz and \Dzb mesons.
The background parameterisation is free to vary independently in each category, whereas the signal shape is shared between WS and RS categories for each \Dstarp flavour.
The RS (WS) yield estimated from the fit corresponds to $11.4\times10^{6}$ ($42,500$) events.

To study the time dependence of the WS/RS ratio, the \deltam fitting procedure is repeated in ten independent \Dz decay-time bins.
Parameters are allowed to differ between bins.
The WS/RS ratio in each bin is calculated from $\sqrt{(N_{\mathrm{WS} \Dz}N_{\mathrm{WS} \Dzb}) / (N_{\mathrm{RS} \Dz}N_{\mathrm{RS} \Dzb})}$, where $N$ denotes the signal yield estimated from the fit for each of the four decay categories.
Using the double ratio ensures that any $\Dstarp/\Dstarm$ production asymmetries or differences in $\slowpion^{+}/\slowpion^{-}$ detection efficiency largely cancel. 

%%%%%%%%%%%%%%%%%%%%%%%%%
%%%%% 5 %%%%%%%%%
%%%%%%%%%%%%%%%%%%%%%%%%%

Several sources of systematic effects are considered that could bias the measured WS/RS ratio.
Candidates in which both a kaon and an oppositely charged pion are misidentified have a very broad structure in $m(\Kp\pim\pip\pim)$, but a signal-like shape in \deltam. 
This background artificially increases the measured WS/RS ratio by causing RS decays to be reconstructed as WS candidates.
In each decay-time bin, $i$, the number of misidentified decays, $N_{\mathrm{ID},i}$, is estimated from WS candidates that are reconstructed further than 40\mevcc from the \Dz mass~\cite{PDG2014}.
The additive correction to the WS/RS ratio is calculated as $\idcori = N_{\mathrm{ID},i} / N_{\mathrm{RS},i}$, where $N_{\mathrm{RS},i}$ is the number of RS decays in the same decay-time bin.
In the entire WS sample it is estimated that $2334\pm65$ misidentified decays are present, constituting $\sim5.5\%$ of the measured WS signal yield.

The decay $\Dz\to\Kp\pim\KS, \KS\to\pip\pim$ has the same final state as signal decays, but a small selection efficiency due to the long flight distance of the \KS. 
Unlike signal decays, the RS and WS categories of this decay have comparable branching fractions~\cite{PDG2014}. 
Assuming that the fraction of $\Dz\to\Km\pip\KS$ decays in the RS sample is negligible, the additive correction to the WS/RS ratio is calculated as, $\kscor = N_{\KS} / N_{RS}$, where $N_{\KS}$ is the number of $\Dz\to\Kp\pim\KS$ decays in the WS sample.
From a fit to both combinations of $m(\pip\pim)$, an estimate of $N_{\KS} = 590 \pm 100$ is obtained, constituting $\sim1.4\%$ of the measured WS signal yield.
This background is observed to have the same decay-time dependence as RS candidates; therefore the same correction of $\kscor = (6.1\pm1.0) \times 10^{-5}$ is applied to the WS/RS ratio in each decay-time bin.

Another background is due to a small fraction of soft pions
that are reconstructed with the wrong charge assignment. 
Such candidates are vetoed by strict requirements on the track quality.
Possible residual background of this type is accounted for by assigning a systematic uncertainty of $2.7\times10^{-5}$ to the measured WS/RS ratio in each decay-time bin.

The systematic uncertainties assigned for $\Dz\to\Kp\pim\KS$ decays and mis-reconstructed soft pions are both expected to be highly correlated between decay-time bins. 
Therefore a correlation coefficient of $1.0$ is used between every pair of decay-time bins, which is confirmed as the most conservative approach.

Additional systematic uncertainties are also included for partially reconstructed decays, which are estimated to make up $\sim 0.25\%$ of the measured WS yield, and the choice of signal and background parameterisations used to determine the signal yields.  
The effect of bin migration due to decay-time resolution has been shown to be negligible~\cite{LHCb-PAPER-2012-038, LHCb-PAPER-2013-053}.

Contributions from secondary decays can bias the measured WS/RS ratio because the \Dz decay time is measured with respect to the PV, which for secondary decays does not coincide with the \Dz production vertex; this causes the \Dz decay time to be overestimated. 
The expected WS/RS ratio in bin $i$ can be written as $\tilde{R_i} \left [ 1 - \seccor \right ]$, where $\tilde{R_i}$ is the expected ratio from prompt \D mesons (those produced at the PV), and \seccor is the correction due to secondary decays.
By measuring the fraction of secondary decays in RS candidates, \secfrac, one can bound \seccor on both sides,
\begin{align}
\secfrac \left [  1 - \frac{R_{\mathrm{max}}( \hat{t_i} )}{R( \hat{t_i} )}  \right ] \leq \seccor \leq \secfrac \left [  1 - \frac{R_{\mathrm{min}}(\hat{t_i})}{R(\hat{t_i})}  \right ]. \label{eqn:seccor}
\end{align}
The function $R( t )$ is defined in \eqnref{eqn:ratio}, and $\hat{t_i}$ is the average decay-time in decay-time bin $i$.
The expressions $R_{\mathrm{min}}(\hat{t_i})$ and $R_{\mathrm{max}}(\hat{t_i})$ give the minimum and maximum of \eqnref{eqn:ratio} in the decay-time range $[0, \hat{t_i}]$.
To determine the secondary fractions, \secfrac, a discriminating variable based on the \Dz\ impact parameter relative to the PV is fitted with both a prompt and secondary component: the PDF describing the former is determined from signal candidates with decay-times smaller than $0.8\tau$, and the PDF describing the latter is found from a subsample of candidates that are compatible with the decay chain $\B\to\Dstarpm\mu X$.
From these fits the secondary fraction is seen to increase monotonically with decay-time from $(1.6\pm1.1)\%$ to $(6.9\pm0.6)\%$.

The efficiency to trigger, reconstruct, and select a \DzToKpipipiWS candidate depends on its location in the 5-dimensional phase space of the decay.
Since there are differences in the amplitude structure between WS and RS decays, the measured WS/RS ratio can be biased.
The efficiency is therefore determined in 5-dimensional phase space bins using simulated data. 
In each decay-time bin this is used to correct the WS/RS yields taking into account the observed 5-dimensional event distribution. 
The resulting multiplicative correction factors to the WS/RS ratio, \effcori, differ from unity by less than a few percent, and increase (decrease) the ratio at low (high) decay times.

The background-subtracted and efficiency corrected WS/RS ratio measured in the $i^{\mathrm{th}}$ decay-time bin is given by $\tilde{r_i} \equiv  r_i \effcori - \idcori - \kscor$, where $r_i$ is the WS/RS ratio estimated from the \deltam fit. The parameters of interest are determined by minimising the \chisq function,
\begin{align}
\chi^{2} (\tilde{\mathbf{r}}, C | \boldsymbol\theta ) &= \sum_{i,j=1}^{10}  \left [ \tilde{r_i} - \tilde{R_i} \left( \boldsymbol\theta  \right) \left [ 1 - \seccor \right ] \right ] \left [ C^{-1} \right ]_{ij} \left [ \tilde{r_j} - \tilde{R_j} \left( \boldsymbol\theta  \right) \left [ 1 - \seccorj \right ]  \right ] \\
&+ \chi^2_{\mathrm{sec}}\left (\boldsymbol\theta \right)  \left [ + \chi^2_{x,y}\left (\boldsymbol\theta \right) \right ], \notag
\end{align}
where $C$ is the full covariance matrix of the measurements, including statistical and systematic uncertainties. Here $\tilde{R_i}\left( \boldsymbol\theta  \right)$ gives the theoretical ratio of WS to RS decay rates (\eqnref{eqn:ratio}), integrated over the $i^{\mathrm{th}}$ decay-time bin, which depends on the fit parameter vector $\boldsymbol\theta = \{ \rKpipipi, \RKpipipi\cdot\yprimekpipipi, \frac{1}{4} (x^2 + y^2) \}$. 
Also included in the determination of $\tilde{R_i}\left( \boldsymbol\theta  \right)$ is the decay-time acceptance, which is found from the RS candidates assuming that their decay-time dependence is exponential. 
The parameters \seccor are free to float in the fit with a Gaussian constraint $\chi^2_{\mathrm{sec}}$.
The mean and width of the Gaussian constraints are defined to be the mid-point and half the difference between the limits in \eqnref{eqn:seccor}, respectively, which are dynamically updated during the fit.
The parameters \secfrac (which are required to calculate these limits) are also Gaussian-constrained to their measured values. 
An alternate fit is also performed where the mixing parameters $x$ and $y$ are constrained to world average values~\cite{HFAG}, $x = (0.371\pm0.158)\times10^{-2}$ and $y = (0.656\pm0.080)\times10^{-2}$ with a correlation coefficient of $-0.361$. 
In this case an additional term, $\chi^2_{x,y}$, is included in the fit and $\boldsymbol\theta = \{ \rKpipipi, \RKpipipi\cdot\yprimekpipipi, x, y \}$. 
The two fit configurations are referred to as `unconstrained' and `mixing-constrained'.

%%%%%%%%%%%%%%%%%%%%%%%%%
%%%%% 6 %%%%%%%%%
%%%%%%%%%%%%%%%%%%%%%%%%%

\begin{figure}[tb]
  \begin{center}
    \includegraphics[width=0.49\linewidth]{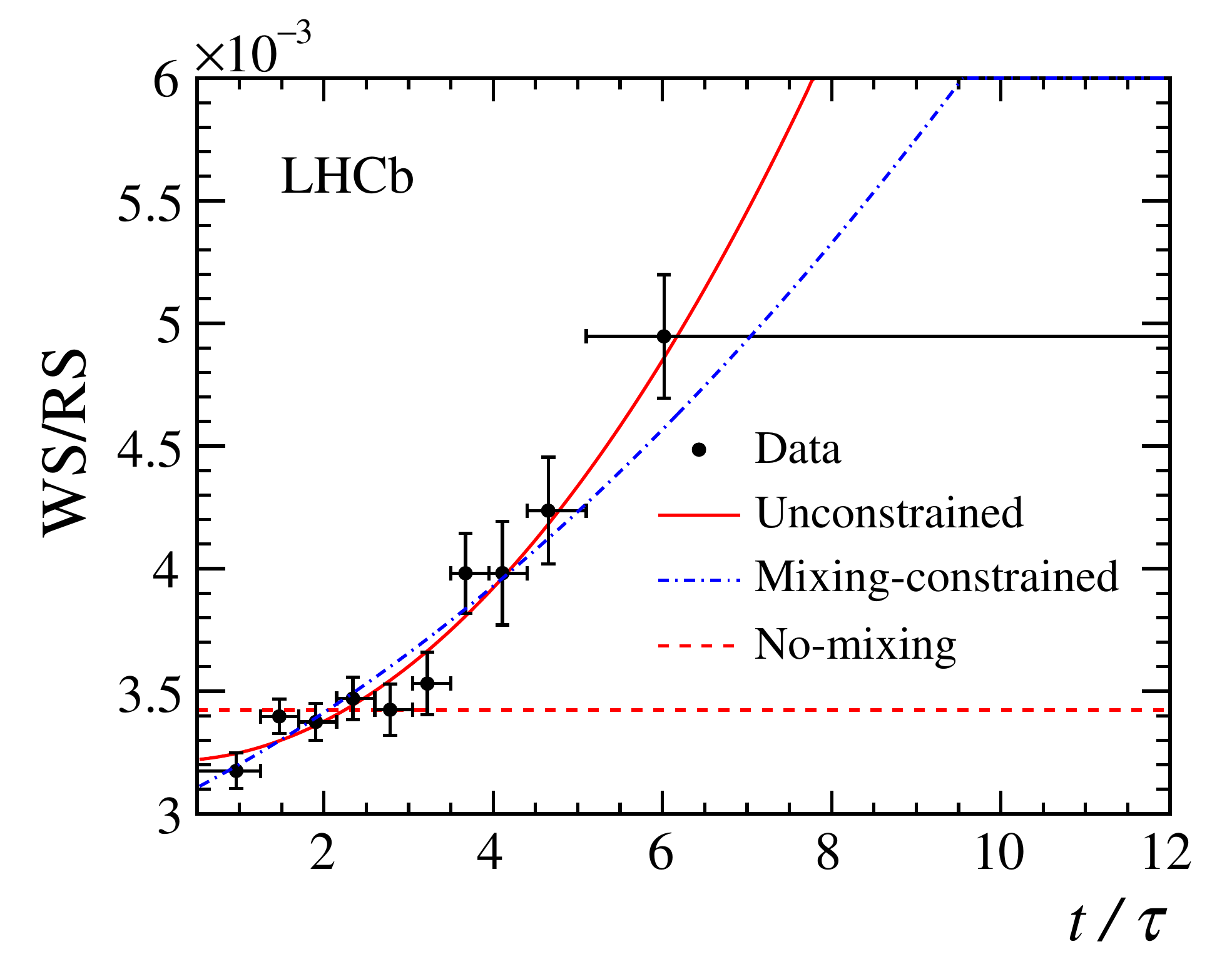}
    \vspace*{-0.7cm}
  \end{center}
  \caption{ Decay-time evolution of the background-subtracted and efficiency corrected WS/RS ratio (points) with the results of the unconstrained (solid line), mixing-constrained (dashed/dotted line), and no-mixing (dashed line) fits superimposed. The bin centres are set to the decay-time where $R(t)$ is equal to the bin integrated ratio $\tilde{R}$ from the unconstrained fit. }
  \label{fig:results}
\end{figure}

\begin{figure}[tb]
  \begin{center}
    \includegraphics[width=0.49\linewidth]{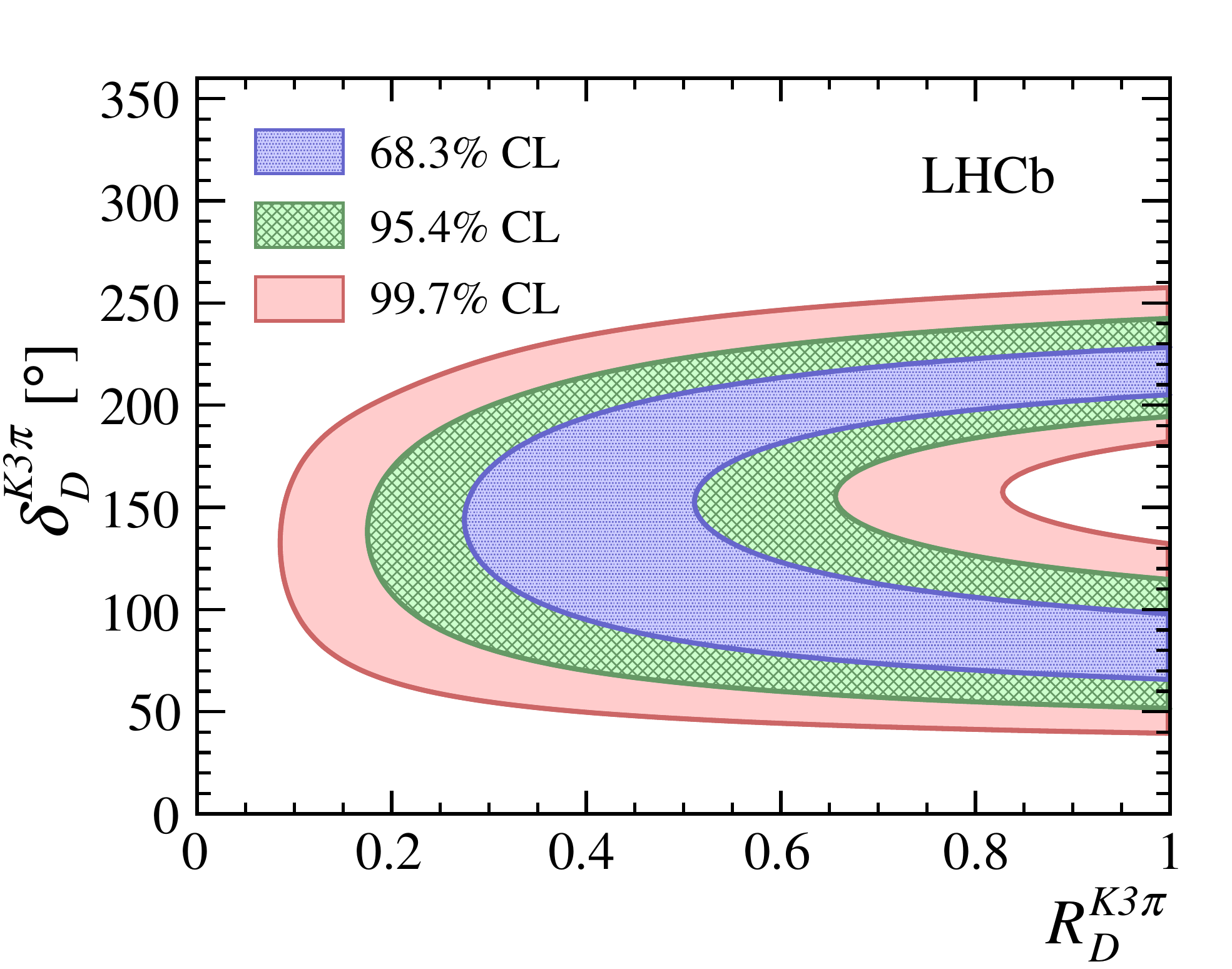}
    \vspace*{-0.7cm}
  \end{center}
  \caption{  Confidence-level (CL) regions in the $\RKpipipi - \delKpipipi$ plane taken from the mixing-constrained fit.  }
  \label{fig:resultsagain}
\end{figure}

\Figref{fig:results} shows the decay-time dependent fits to the WS/RS ratio for the unconstrained, mixing-constrained, and no-mixing fit configurations; the latter has the fit parameters $\RKpipipi\cdot\yprimekpipipi$ and $\frac{1}{4} (x^2 + y^2)$ fixed to zero.  
The numerical results of the unconstrained and mixing-constrained fit configurations are presented in \tabref{tab:results}.
The values of $\RKpipipi\cdot\yprimekpipipi$ and $\frac{1}{4} (x^2 + y^2)$ from the unconstrained fit are both compatible with zero at less than 3 standard deviations, but due to the large correlation between these parameters, the hypothesis
that both are zero can be rejected with much higher significance.
Using Wilks' theorem~\cite{wilks1938} the no-mixing hypothesis is excluded at a significance level of 8.2 standard deviations. 
The value of $\frac{1}{4} (x^2 + y^2)$ determined using the world average values of $x$ and $y$ is compatible with the unconstrained fit result at 1.8 standard deviations.
The results of the mixing-constrained fit show that the uncertainties
on the parameters \rKpipipi and $\RKpipipi\cdot\yprimekpipipi$ are
reduced by $41\%$ and $61\%$ respectively in comparison with the
unconstrained fit.  Using the mixing-constrained fit, it is possible
to identify a line of solutions in the $(\RKpipipi, \delKpipipi)$
plane. The two-dimensional contours containing $68.3\%$, $95.4\%$ and
$99.7\%$ confidence regions are shown in \figref{fig:resultsagain}.
The only other constraints on $(\RKpipipi, \delKpipipi)$ are based on
\cleoc data~\cite{Libby:2014rea}.  
A combination would require a
combined fit sharing the input on $x$ and $y$. 
A combination made ignoring this complication shows that the input from mixing results in reductions
in uncertainties on $\RKpipipi$ and $\delKpipipi$ by approximately 50\% when compared to the \cleoc values.

\begin{table}[t]
  \caption{ Results of the decay-time dependent fits to the WS/RS ratio for the unconstrained and mixing-constrained fit configurations. The results include all systematic uncertainties. }
\begin{center}\begin{tabular}{lcrcccc}
    \hline \hline
    Fit Type                           & Parameter & \multicolumn{1}{c}{Fit result}   & \multicolumn{4}{c}{Correlation coefficient}  \\
    $\chi^2$/ndf (p-value)                  &                  &    & $\rKpipipi$ & $\RKpipipi\cdot\yprimekpipipi$ & \multicolumn{2}{c}{$\frac{1}{4} (x^2 + y^2)$}  \\ [0.5ex]       
    \hline \noalign{\vskip 0.7mm}
    Unconstrained & $\rKpipipi$                       & $(5.67 \pm 0.12)\times 10^{-2}$                                       & 1 & 0.91 & \multicolumn{2}{c}{0.80} \\
     $7.8/7 \ (0.35)$            & $\RKpipipi\cdot\yprimekpipipi$                &   $\phantom{0}(0.3  \pm 1.8)\phantom{0}\times 10^{-3}$  &  & 1 & \multicolumn{2}{c}{0.94} \\
                            & $\frac{1}{4} (x^2 + y^2)$ &   $\phantom{0}(4.8 \pm 1.8)\phantom{0}\times 10^{-5}$   &  &  &  \multicolumn{2}{c}{1} \\ [0.5ex]   
    \hline \noalign{\vskip 0.7mm}
      & & & $\rKpipipi$ & $\RKpipipi\cdot\yprimekpipipi$ & $x$ &  $y$ \\[0.5ex]   
    \hline \noalign{\vskip 0.7mm}
     Mixing-constrained & $\rKpipipi$       & $(5.50 \pm 0.07)\times 10^{-2}$ & 1 & 0.83 & 0.17 & 0.10 \\
       $11.2/8 \ (0.19)$                   & $\RKpipipi\cdot\yprimekpipipi$ & $(-3.0 \pm 0.7)\phantom{0}\times 10^{-3}$ & & 1 & 0.34 & 0.20 \\   
                                    & $x$                   & $\phantom{0}(4.1  \pm 1.7)\phantom{0}\times 10^{-3}$ & & & 1 & -0.40 \\   
                                     & $y$                  & $\phantom{0}(6.7 \pm 0.8)\phantom{0}\times 10^{-3}$ & & &  & 1 \\   
   \noalign{\vskip 0.7mm} \hline
    \hline
  \end{tabular}\end{center}
\label{tab:results}
\end{table}

To evaluate the impact of systematic uncertainties included in the result, the fits are repeated with the systematic uncertainties on the WS/RS ratio set to zero.
In the unconstrained fit the uncertainties in $\rKpipipi$, $\RKpipipi\cdot\yprimekpipipi$ and $\frac{1}{4} (x^2 + y^2)$ are reduced by $11\%$, $9\%$ and $11\%$, respectively. 
In the mixing-constrained fit the uncertainties in $\rKpipipi$ and $\RKpipipi\cdot\yprimekpipipi$ are reduced by $15\%$ and $9\%$, respectively.

Using the results presented in \tabref{tab:results} the decay-time integrated WS/RS ratio, $R^{K3\pi}_{\mathrm{WS}} = (\rKpipipi)^2 - \rKpipipi \RKpipipi\cdot\yprimekpipipi + \frac{1}{2}(x^2+y^2)$, is calculated to be $(3.29\pm0.08) \times 10^{-3}$ for the unconstrained result, and $(3.22\pm0.05) \times 10^{-3}$ for the mixing-constrained result. 
This is consistent with the existing measurement from Belle~\cite{BelleK3piUpdate}, and has smaller uncertainties.
Using the RS branching fraction, $\mathcal{B}(\DzToKpipipiRS) = (8.07\pm0.23) \times 10^{-2}$~\cite{PDG2014}, the WS branching fraction, $\mathcal{B}(\DzToKpipipiWS)$, is determined to be $(2.66\pm0.06\pm0.08)\times10^{-4}$ using the unconstrained result, and $(2.60\pm0.04\pm0.07)\times10^{-4}$ using the mixing-constrained result. 
Here the first uncertainty is propagated from $R^{K3\pi}_{\mathrm{WS}}$ and includes systematic effects, and the second is from the knowledge of $\mathcal{B}(\DzToKpipipiRS)$. 

%%%%%%%%%%%%%%%%%%%%%%%%%
%%%%% 7 %%%%%%%%%
%%%%%%%%%%%%%%%%%%%%%%%%%

In conclusion, the decay-time dependence of the ratio of \DzToKpipipiWS to \DzToKpipipiRS decay rates is observed, and the no-mixing hypothesis is excluded at a significance level of $8.2$ standard deviations.
The worldÕs most precise measurements of $\rKpipipi$ and $R^{K3\pi}_{\mathrm{WS}}$ are presented, and a unique constraint on $\RKpipipi\cdot\yprimekpipipi$ is given, which will increase sensitivity to the CP-violating phase \gam in \BpToDKp, \DToKmpippimpip decays.

\section*{Acknowledgements}

\noindent We express our gratitude to our colleagues in the CERN
accelerator departments for the excellent performance of the LHC. We
thank the technical and administrative staff at the LHCb
institutes. We acknowledge support from CERN and from the national
agencies: CAPES, CNPq, FAPERJ and FINEP (Brazil); NSFC (China);
CNRS/IN2P3 (France); BMBF, DFG and MPG (Germany); INFN (Italy); 
FOM and NWO (The Netherlands); MNiSW and NCN (Poland); MEN/IFA (Romania); 
MinES and FANO (Russia); MinECo (Spain); SNSF and SER (Switzerland); 
NASU (Ukraine); STFC (United Kingdom); NSF (USA).
We acknowledge the computing resources that are provided by CERN, IN2P3 (France), KIT and DESY (Germany), INFN (Italy), SURF (The Netherlands), PIC (Spain), GridPP (United Kingdom), RRCKI and Yandex LLC (Russia), CSCS (Switzerland), IFIN-HH (Romania), CBPF (Brazil), PL-GRID (Poland) and OSC (USA). We are indebted to the communities behind the multiple open 
source software packages on which we depend.
Individual groups or members have received support from AvH Foundation (Germany),
EPLANET, Marie Sk\l{}odowska-Curie Actions and ERC (European Union), 
Conseil G\'{e}n\'{e}ral de Haute-Savoie, Labex ENIGMASS and OCEVU, 
R\'{e}gion Auvergne (France), RFBR and Yandex LLC (Russia), GVA, XuntaGal and GENCAT (Spain), The Royal Society, Royal Commission for the Exhibition of 1851 and the Leverhulme Trust (United Kingdom).

\clearpage

\addcontentsline{toc}{section}{References}
\setboolean{inbibliography}{true}
\bibliographystyle{LHCb}
\ifx\mcitethebibliography\mciteundefinedmacro
\PackageError{LHCb.bst}{mciteplus.sty has not been loaded}
{This bibstyle requires the use of the mciteplus package.}\fi
\providecommand{\href}[2]{#2}

\newpage

%%%%%%%%%%%%%%%%%%%%%%%%%%%%%%%%%%%%%%%%%%
\centerline{\large\bf LHCb collaboration}
\begin{flushleft}
\small
R.~Aaij$^{39}$, 
C.~Abell\'{a}n~Beteta$^{41}$, 
B.~Adeva$^{38}$, 
M.~Adinolfi$^{47}$, 
A.~Affolder$^{53}$, 
Z.~Ajaltouni$^{5}$, 
S.~Akar$^{6}$, 
J.~Albrecht$^{10}$, 
F.~Alessio$^{39}$, 
M.~Alexander$^{52}$
% ...

{\footnotesize \it
% ...
$ ^{5}$Clermont Universit\'{e}, Universit\'{e} Blaise Pascal, CNRS/IN2P3, LPC, Clermont-Ferrand, France\\
$ ^{6}$CPPM, Aix-Marseille Universit\'{e}, CNRS/IN2P3, Marseille, France\\
% ...
$ ^{10}$Fakult\"{a}t Physik, Technische Universit\"{a}t Dortmund, Dortmund, Germany\\
% ...
$ ^{38}$Universidad de Santiago de Compostela, Santiago de Compostela, Spain\\
$ ^{39}$European Organization for Nuclear Research (CERN), Geneva, Switzerland\\
% ...
$ ^{41}$Physik-Institut, Universit\"{a}t Z\"{u}rich, Z\"{u}rich, Switzerland\\
% ...
$ ^{47}$H.H. Wills Physics Laboratory, University of Bristol, Bristol, United Kingdom\\
% ...
$ ^{52}$School of Physics and Astronomy, University of Glasgow, Glasgow, United Kingdom\\
$ ^{53}$Oliver Lodge Laboratory, University of Liverpool, Liverpool, United Kingdom\\
% ...
}
\end{flushleft}
%%%%%%%%%%%%%%%%%%%%%%%%%%%%%%%%%%%%%%%%%%

\newpage

\end{document}